# Continuous Global Optimization in Surface Reconstruction from an Oriented Point Cloud

Rongjiang Pan[1], Vaclav Skala[2]

[1]School of Computer Science and Technology, Shandong University, Jinan, China,
panrj@sdu.edu.cn

[2]Centre of Computer Graphics and Data Visualization, Department of Computer Science and Engineering, University of West Bohemia, Plzen, Czech Republic
http://Graphics.zcu.cz

## Abstract

*We introduce a continuous global optimization method to the field of surface reconstruction from discrete noisy cloud of points with weak information on orientation. The proposed method uses an energy functional combining flux-based data-fit measures and a regularization term. A continuous convex relaxation scheme assures the global minima of the geometric surface functional. The reconstructed surface is implicitly represented by the binary segmentation of vertices of a 3D uniform grid and a triangulated surface can be obtained by extracting an appropriate isosurface. Unlike the discrete graph-cut solution, the continuous global optimization entails advantages like memory requirements, reduction of metrication errors for geometric quantities, allowing globally optimal surface reconstruction at higher grid resolutions. We demonstrate the performance of the proposed method on several oriented point clouds captured by laser scanners. Experimental results confirm that our approach is robust to noise, large holes and non-uniform sampling density under the condition of very coarse orientation information.*

## 1 Introduction

Point clouds produced by 3D scanners are not directly used in many practical applications. Reconstructing watertight surfaces from point samples is becoming a common step in modeling real-world objects. The reconstruction problem has been researched extensively and many techniques have been developed in the last two decades [12,17]. However, surface reconstruction still remains a difficult problem and an active research area because the scanned data can be non-uniform, contaminated by some noise. Moreover, inaccessibility during scanning and some material properties may leave portions of the surface devoid of point samples. Additionally, a lack of accurate information about the surface orientation at the point samples is known to be a main challenge common with surface reconstruction methods using an implicit function framework.

To deal with most of the difficulties with surface reconstruction from cloud of points, we formulated surface reconstruction as an energy minimization problem by combining flux of a sparse normal vector field with a minimal surface regularization term, similarly to the previous work of [16]. Flux maximizing flow is first introduced into shape optimization problem by Vasilevskiy and Siddiqi [29]. Minimal surface regularization methods usually yield watertight surfaces and are robust to noise as well. Unfortunately, the resulting minimization problem is non-convex, much sensitive to initialization and may have more local minima, making the optimization method a crucial issue to the quality of the reconstruction.

The main contribution of the presented work is the development of a continuous global optimization technique for the energy minimization problem, which reconstructs a globally optimal surface from noisy cloud of points with only weak orientation information. This approach was inspired by the works in the context of image segmentation [9] and multiview 3D reconstruction [20]. Unlike the discrete graph-cut solution [16], the continuous global optimization entails advantages like memory requirements, reduction of metrication errors [19] for geometric quantities, allowing globally optimal surface reconstruction at higher grid resolutions.

We also show that the energy model leads to a Poisson equation by minimizing the square of the variation of the embedding implicit function instead of the minimal surface regularization term. Hence, the Poisson surface reconstruction proposed in [15] can be regarded as a special case of our proposed method.

The paper is organized as follows. The next section contains a brief review of some related approaches. In Section 3, we present the underlying theory, i.e. energy model and the optimization of the functional.

Section 4 provides the implementation details. In Section 5, we show some experimental results and future work is discussed in Section 6.

## 2 Related Work

Many techniques have been developed to reconstruct a surface from discrete point samples of the object surface. We can broadly classify most of previous works into combinational structure approaches and volumetric reconstruction techniques.

Approaches based computational geometry, such as Delaunay triangulations [7], alpha shapes [5,6] and Voronoi diagrams [3,11], create triangular mesh which interpolates all or most of the points. Although with the strength of reconstructing fine surface details, it is often difficult to get a smooth surface in the presence of noise and inhomogeneous sample density.

Volumetric reconstruction methods generally fit an implicit function to the point samples, and then represent the reconstructed surface as an appropriate isosurface of this function. Relevant approaches are based on signed distance functions [10,12], radial basis functions [8,21,22,25,28], local implicit functions [23], moving least square approximation [1,4] and indicator functions [14,15]. Implicit surface methods are usually robust to noise and non-uniform sampling density. Complicated topology and Boolean operations are easily handled with these methods. However, the generation of the implicit function relies on an ability to distinguish between the inside and outside of the closed surface. This is known to be the main challenge in the reconstruction pipeline. These methods usually require additional information, such as a sample's normal estimation from its neighbors [12], classified poles of the Voronoi diagram of the input points [25], heuristically computed inside/outside constraints [24]. In the presence of noise or thin features, this additional information is highly unreliable and often leads to an erroneous surface reconstruction.

Other volumetric approaches try to reconstruct a surface approximation from unoriented point sets [2,13,14].Without orientation information, however, these algorithms may lead to over-smoothing surface [13] and they cannot deal with large gaps [2] or large data sets [14].

Recently, graph-cut optimization algorithms were adapted to surface fitting problem [16]. With a flux-based functional, the global optimization method reconstructs watertight surfaces in presence of noise, outliers, large missing parts and orientation errors at the data points. Nevertheless, the computation and memory requirement quickly becomes prohibitively expensive for higher grid resolutions even with automatically adjusted sub-graphs.

There is a continuous global optimization technique in the context of image segmentation [9] and multiview 3D reconstruction [20], which allows to avoid some limitations of the graph cuts approach. The continuous optimization method entails several advantages such as an absence of metrication errors, lower memory requirements. Detailed comparison between discrete and continuous optimization methods can be found in [19].

Our work extends the power of the continuous global optimization technique to surface reconstruction from oriented point clouds.

## 3 The energy model

Let $S$ be an input data set of samples lying on or near the surface $\partial M$ of an unknown model $M$. Each sample $s \in S$ consists of a point $s.p$ and a weakly estimated outward-pointing normal $s.\mathbf{v}$. Our goal is to compute the indicator function $u$ (defined as 1 at points inside the model and 0 at points outside) of the model $M$, and then to approximate the surface $\partial M$ by a watertight, triangulated isosurface.

We consider an energy model combining geometric functional with regularization. Let $V \in R^3$ be a volumetric region enclosing the model $M$, and $\mathbf{v}:V \to R^3$ a vector field estimated from normals at the input data points, whose restriction to surface $\partial M$ represents an estimate of the outward orientation of surface $\partial M$. We consider the following energy minimization problem:

$$\min \ E(u) = \lambda \int_V \phi(|\nabla u|)dx + \int_{\partial M} <\mathbf{v}(x), \nabla u> dx \qquad (1)$$
$$\text{s.t.} \ u \in \{0,1\}$$

where $<\mathbf{v}(x), \nabla u>$ denotes the dot product between the vector filed $\mathbf{v}$ and the gradient field $\nabla u$. The first term in $E(u)$ is a smoothing constraint. The second item measures the consistency between the estimated vector field $\mathbf{v}$ and gradient field $\nabla u$ on surface $\partial M$. We wish to compute the indicator function $u$ such that its gradient is best aligned (in opposite directions) with the normal field. The idea has been developed in Hermite variational implicit surface reconstruction [31] and is similar to the anisotropic metric for multiview reconstruction developed independently in [30]. The parameter $\lambda > 0$ determines in some sense the smallest feature that will be maintained in the reconstructed surface. The function $\phi : R^+ \rightarrow R^+$ is to be defined as a function of the first-order variation $|\nabla u|$.

After changing the implicit surface orientation to outward to apply the divergence theorem, i.e. the volume integral of the divergence $\text{div}(\mathbf{v})$ of continuously differentiable vector field $\mathbf{v}$ over volume $M$ equals the surface integral of $\mathbf{v}$ over the boundary $\partial M$ of the volume $M$, we can obtain the following optimization problem:

$$\min \ E(u) = \lambda \int_V \phi(|\nabla u|) dx - \int_{interior(M)} \text{div}(\mathbf{v}) dx \tag{2}$$
$$\text{s.t.} \ u \in \{0,1\}$$

where $interior(M)$ denotes the interior of model $M$.

With the implicit representation $u : V \rightarrow \{0,1\}$ of model $M$, we have the following constrained, nonconvex energy minimization problem corresponding to (2):

$$\min \ E(u) = \lambda \int_V \phi(|\nabla u|) dx - \int_V \text{div}(\mathbf{v}) u(x) dx \tag{3}$$
$$\text{s.t.} \ u \in \{0,1\}$$

Since the space of binary functions is a non-convex space, the energy optimization problem in (3) is also non-convex. However, relaxing the binary condition to $u \in [0,1]$ yields a constrained convex optimization problem,

$$\min \ E(u) = \lambda \int_V \phi(|\nabla u|) dx - \int_V \text{div}(\mathbf{v}) u(x) dx \tag{4}$$
$$\text{s.t.} \ u \in [0,1]$$

which is addressed in several works [9,20]. The global minimizers of (4) are not unique in general since it is not strictly convex; however, it does not have any local minima that are not global minima. Moreover, the above energy functional has a very nice property that allows global minimization of the original 3D segmentation problem (1) as stated by the following theorem. The proof is very similar to [9,20] and here we provide only the result.

Theorem 1: If $u^* : V \rightarrow [0,1]$ is any minimizer of the functional (4), then for almost every threshold $\mu \in (0,1)$ the binary function

$$u(x) = \begin{cases} 1 & u^*(x) > \mu \\ 0 & otherwise \end{cases}$$

is also a minimizer of (1).

The minimum of (4) must satisfy the Euler-Lagrange equation

$$0 = -\lambda \text{div}\left( \frac{\phi'(|\nabla u|)}{|\nabla u|} \nabla u \right) - \text{div}(\mathbf{v}), \ u \in [0,1] \tag{5}$$

The function $\phi(|\nabla u|) = |\nabla u|^2$, namely, the $L^2$ norm of the gradient of $u$, was proposed in [27]. Then the Euler-Lagrange equation would be:

$$0 = -2\lambda \Delta u - \text{div}(\mathbf{v}), \ u \in [0,1] \tag{6}$$

Thus, the variation problem (4) transforms into a standard Poisson problem. Since Laplacian operator has very strong isotropic smoothing properties and does not preserve edges, the reconstructed surface is comparatively over smoothed.

Most of the works [9,16,20] proposed to use the $L^1$ norm of the gradient of $u$, also called the total variation. It is provable that the area of the surface is also minimized by minimizing the total variation of the characteristic function $u$ [9]. In this case, (5) becomes

$$0 = -\lambda \text{div}\left( \frac{\nabla u}{|\nabla u|} \right) - \text{div}(\mathbf{v}), \ u \in [0,1] \tag{7}$$

## 4   Implementation

We now give the implementation details of the proposed approach.

### 4.1 Vector field definition

Since the discretization of gradient and divergence operators is simple and more accurate on a uniform grid, we first discretize the problem on a regular 3D grid, as shown in Figure 1.

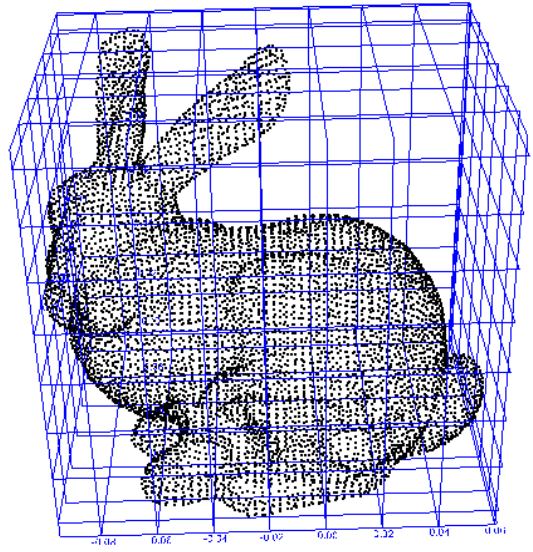

Figure 1: Regular 3D grid discretizing the domain of the input points

Instead of estimating a dense globally oriented vector field by some heuristic, which is generally regarded as a computationally intensive and conceptually difficult problem especially in the presence of noise and thin features [13], we assumed a weak estimate of global surface orientation $s.\mathbf{v}$ at each sample $s \in S$. Like [16], the experimental results in this paper use the direction from a sample to its corresponding sensor/camera, except the samples come with orientation information.

The weak estimate $s.\mathbf{v}$ at each sample $s \in S$ is distributed to its eight nearest grid vertices as follows $s.\mathbf{v}\cdot(1-x)(1-y)(1-z)$, $s.\mathbf{v}\cdot x(1-y)(1-z)$, $s.\mathbf{v}\cdot(1-x)y(1-z)$, $s.\mathbf{v}\cdot(1-x)(1-y)z$, $s.\mathbf{v}\cdot x(1-y)z$, $s.\mathbf{v}\cdot(1-x)yz$, $s.\mathbf{v}\cdot xy(1-z)$, $s.\mathbf{v}\cdot xyz$, where $x$, $y$ and $z$ are the **differences** between the coordinates of point $s.p$ and the smallest coordinates among the eight vertices with grid spacing 1. In order to approximate a dense vector field $\{\mathbf{v}(x)|x \in V\}$, we smooth the sparse vector field on the regular 3D grid with a Gaussian. In practice, we use the $n$-th convolution of a box filter with itself:

$$B(t) = \begin{cases} 1 & |t| < h \\ 0 & otherwise \end{cases}$$

where $h$ is the size of grid cell and we choose $n = 3$ in our implementation. Then, $\text{div}(\mathbf{v}) = \frac{\partial v_1}{\partial x_1} + \frac{\partial v_2}{\partial x_2} + \frac{\partial v_3}{\partial x_3}$ is approximated by standard central differences, where $\mathbf{v} = (v_1, v_2, v_3)^{\text{T}}$.

### 4.2 Numerics

As the explicit gradient descent scheme converges very slowly, we use the successive over-relaxation (SOR) method proposed in [20] to solve the nonlinear system (7).

Starting with an initialization $u^0 = 0$, we compute the diffusivity $g = \frac{1}{\sqrt{|\nabla u|^2 + \varepsilon^2}}$ and keep it constant; $\varepsilon$ is a small constant that avoids infinite diffusivity when $|\nabla u| = 0$.

The Eq.(7) is linear when constant $g$ is fixed and can be solved with a fixed point iteration scheme. For each grid vertex $i$, we update $u_i$ by

$$u_i^{k+1} = (1-\omega)u_i^k + \omega \frac{\lambda \sum_{j \in N(i), j<i} g_{ij} u_j^{k+1} + \lambda \sum_{j \in N(i), j>i} g_{ij} u_j^k + div_i}{\lambda \sum_{j \in N(i)} g_{ij}} \qquad (8)$$

where $u_i^k$ denotes $u_i$ at iteration $k$, $N(i)$ contains the 6-neighborhood of grid vertex $i$ and $g_{ij} = (g_i + g_j)/2$ represents the diffusivity between grid vertex $i$ and its neighbor $j$. The optimal relaxation parameter $\omega$ depends on the linear system and has to chosen between 0 and 2 for the method to converge. In our implementation, we set $\varepsilon = 0.001$, $\omega = 1.85$ according to [20] and update the diffusivity $g$ every two iterations. Iterations are stopped when the relative error of the energy in two successive iterations is around the machine accuracy or reach the given limit by a user.

### 4.3 Multi-resolution

To reduce the computation overhead, we also applied a multi-resolution scheme in the proposed method. In our implementation, we use a three-level pyramid. The $div(\mathbf{v})$ is computed at finest grid only once, and then down-sampled by summation. The result computed at lower resolution is up-sampled as an initialization of the higher level. Therefore, the optimization may concentrate on a very small subset of the domain of interest.

### 4.4 Isosurface extraction

Based on the Theorem 1, in order to obtain a binary characteristic function $\hat{u}$ for the model $M$, we simply threshold the solution of the convex problem with $\mu \in (0,1)$. In our experiments, we chose $\mu = 0.5$, however, the difference between the results with $\mu \in [0.1, 0.9]$ was quite small. For the Stanford Bunny at grid resolution $212 \times 200 \times 159$, table 1 shows the number of grid vertices whose $\hat{u}$ is 1 with several different threshold values of $\mu$.

| threshold $\mu$ | 0.1 | 0.25 | 0.5 | 0.75 | 0.9 |
|---|---|---|---|---|---|
| number of grid vertices inside the model | 1,126,101 | 1,124,452 | 1,121,936 | 1,119,334 | 1,117,737 |

Table 1: Stats for different threshold values with the Stanford Bunny.

In order to reconstruct a triangulated surface $\partial M$, it is necessary to select an isovalue and extract the corresponding isosurface from the binary characteristic function $\hat{u}$. For 3D rendering, however, such isosurfaces exhibit distracting aliasing artifacts. We used a smoothed version $\tilde{u}$ of the binary characteristic function $\hat{u}$ by a similar smoothing filter in Section 4.1.

The isovalue $\mu$ is selected as the weighted average of the values of $\tilde{u}$ at the sample positions:

$$\mu = \frac{1}{|S|} \sum_{s \in S} \tilde{u}(s.p)$$

where $\tilde{u}(s.p)$ denotes the trilinear interpolation to the eight nearest grid vertices of $s$. Finally, we use an adaptation of the Marching Cubes code [18] to extract the isosurface.

## 5  Results

We validated our approach on a series of experiments. The proposed method was implemented in C++ on a notebook with 2.26GHz Core 2 Duo CPU and 2GB of RAM.

We first performed experiments on laser-scanned data sets downloaded from the Stanford 3D Scanning Repository [26]. The registered raw range scans were our input and a single orientation vector corresponding to scan viewing direction was assigned to all points in the same scan.

Figure 2 illustrates our reconstruction results for the Dragon model with different $\lambda$ values. The parameter $\lambda$ affects the fitness to the sample points and smoothness of the surface. Large values of $\lambda$ lead to increased smoothing. It can be estimated by the user or chosen automatically using k-fold cross validation. In our experiments, we got reasonable results with $\lambda$ between 0.005 and 0.01.

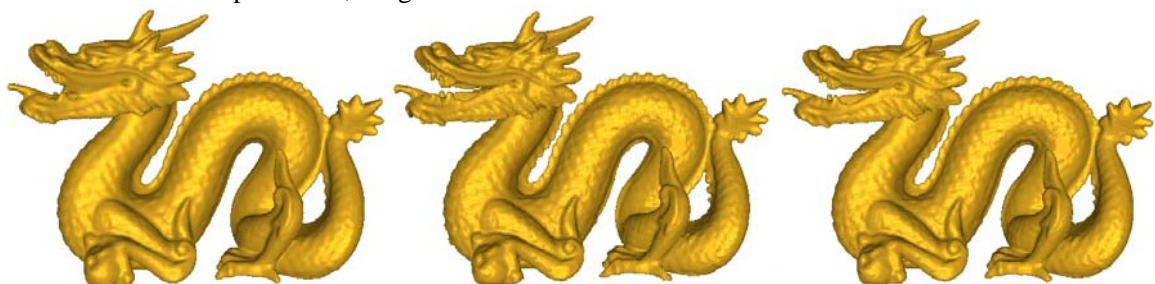

Figure 2: Reconstruction of the Dragon model at various values of $\lambda$ : 0.1 (left), 0.01 (middle), 0.001(right)

For the thin drill bit assembled from 12 range images, the false edge extensions inherent in data from triangulation scanners pose a challenge to surface reconstruction. As Figure 3 demonstrates, our method behaves robustly.

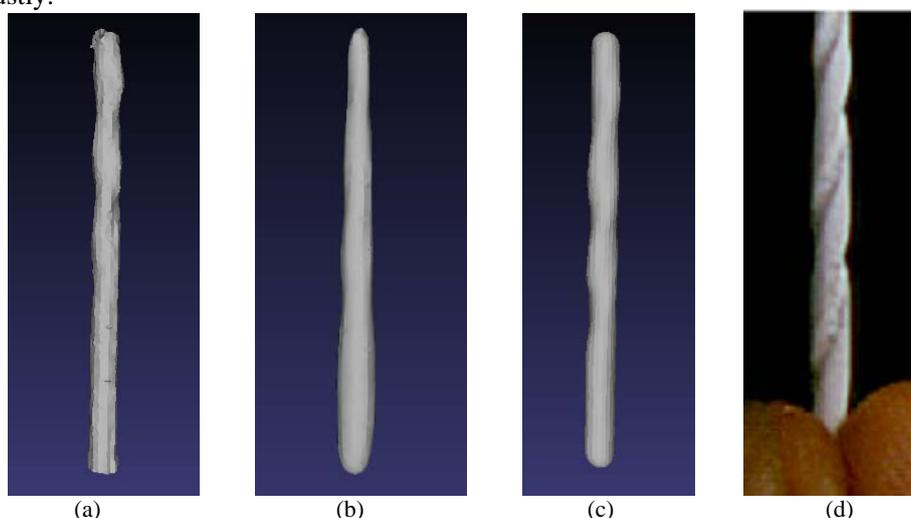

(a)　　　　　(b)　　　　　(c)　　　　　(d)
Figure 3: Reconstructions of the drill bit using VRIP (a), Poisson surface (b) and our method (c). A photograph of the original drill bit [10] is shown in (d).

To study scalability with large variations in sampling density and some outliers, we removed 98% of points from one half of Armadillo and kept the outliers added by scanning process. Unlike [16] which used non-uniform Euclidean regularization, our method was able to handle the 50-to-1 difference in density and tolerate outliers without using any other information, as shown in Figure 4.

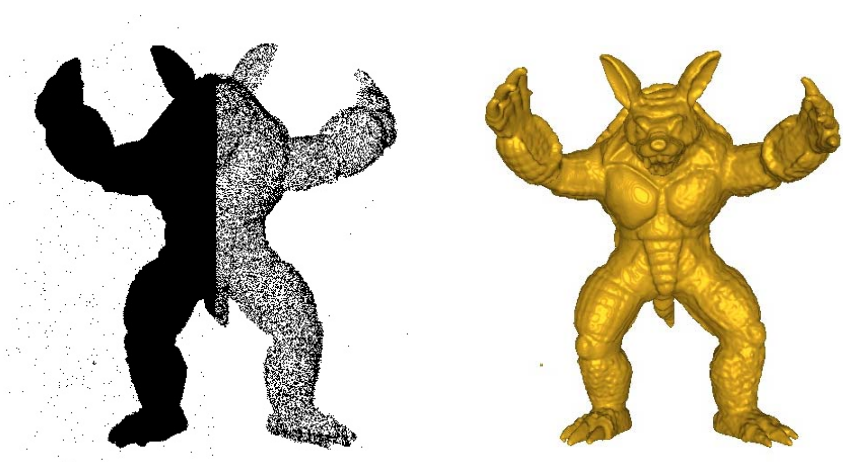

Figure 4: Reconstruction result (right) of the Armadillo range scans with 50-to-1 difference in density (left).

The Buddha model reconstruction proofs other property of the proposed method. Since there are no samples between the two feet of Buddha, light-of-sight information near the legs were used to fill the large hole and disconnect the two feet in [10,16]. In contrast, we correctly reconstruct the regions without using additional information, as shown in Figure 5.

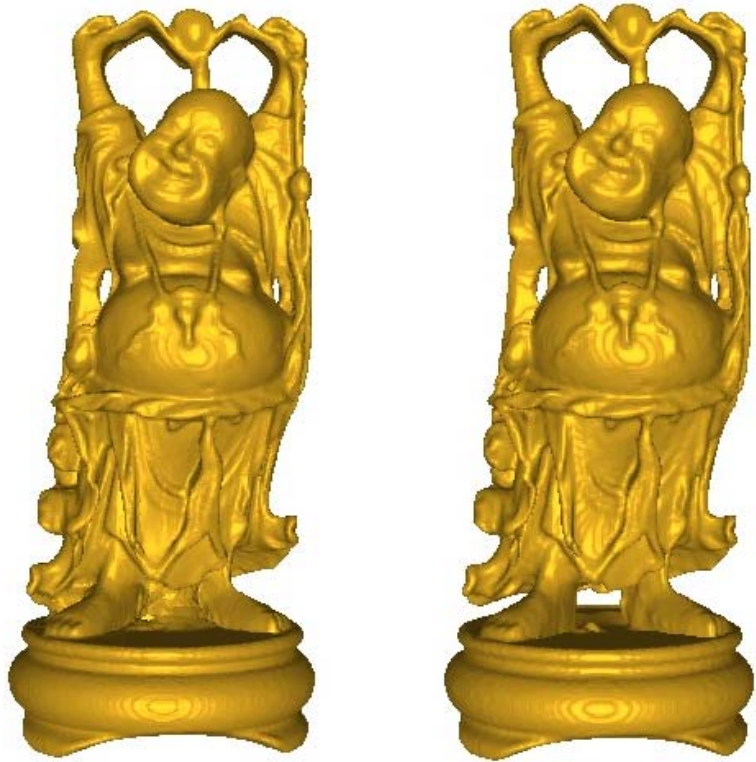

Figure 5: Reconstructions of the Buddha using Poisson surface (left) and our method (right).

Reconstructions at intermediate levels for the Bunny data set are shown in Figure 6.

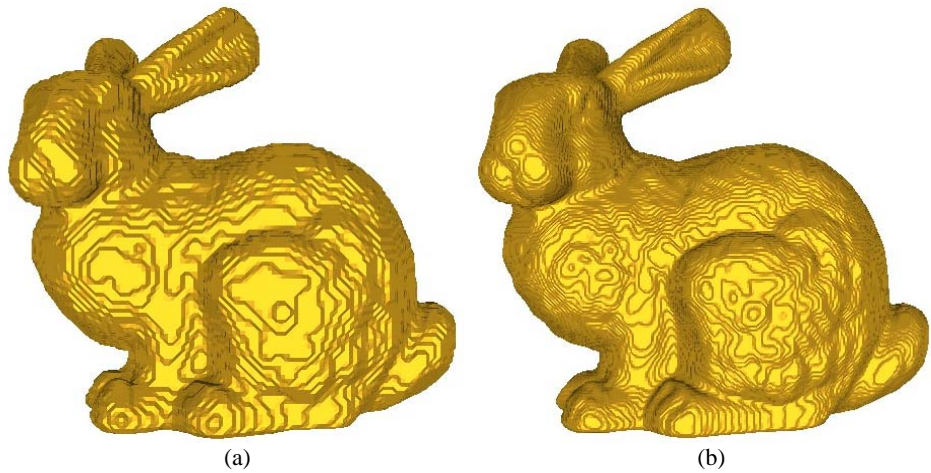

(a)    (b)

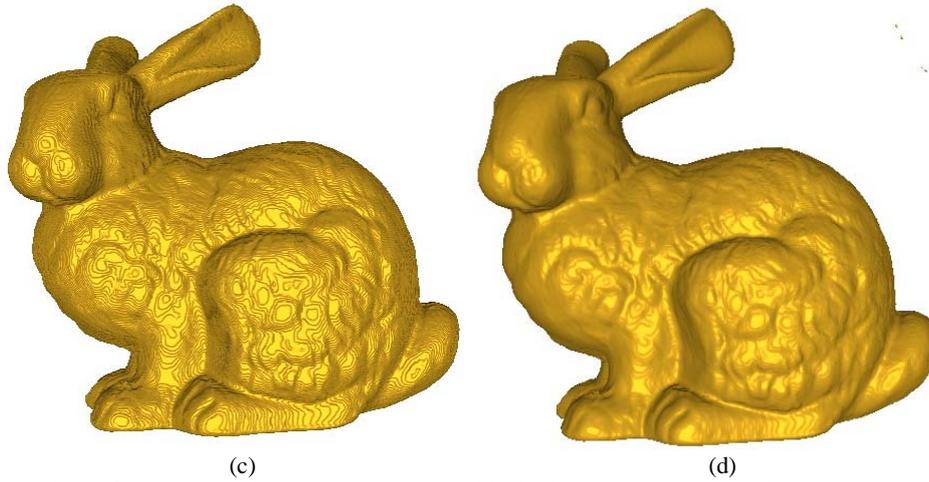

(c)                                      (d)

Figure 6: Reconstructions of the Bunny using a three-level pyramid with increasing grid resolutions: (a)$128\times 120\times 94$, (b)$256\times 241\times 189$, (c)$512\times 482\times 378$. The smoothed surface is shown in (d).

Table 2 summarizes the temporal and spatial performance of our algorithm in the experiments mentioned above. On the contrary, most of the experiments are prohibitive if graph-cut optimization is used due to the higher amount of memory demand.

| model | Grid size | Time [s] | Peak Memory [MB] | Triangles |
|---|---|---|---|---|
| Bunny | $512\times 482\times 378$ | 292 | 719 | 1,378,268 |
| Dragon | $612\times 436\times 286$ | 241 | 585 | 1,855,544 |
| Armadillo | $486\times 512\times 408$ | 345 | 780 | 1,091,624 |
| Buddha | $346\times 812\times 344$ | 452 | 752 | 2,618,518 |

Table 2: Grid size, computation time (in seconds), peak memory usage (in megabytes), and number of triangles in the reconstructed surface.

For 180 points uniformly sampled on a sphere, Figure 7 compares the result of our reconstruction algorithm to the results obtained using graph-cut method [16]. The metrication errors appear in all configuration of graph-cut. The accuracy of the reconstructed surface is estimated by the RMS (root mean square) of the distances from the input points to the nearest points on the surface.

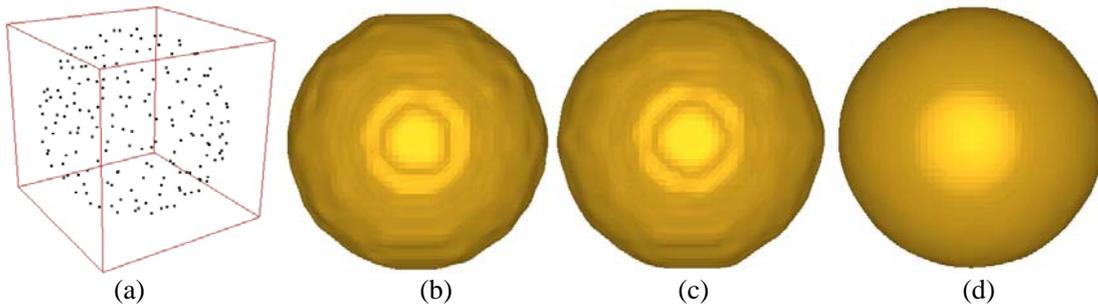

(a)                  (b)                  (c)                  (d)

Figure 7: Reconstructions from 180 points on a sphere (a) using discrete graph-cut algorithm of 6-connected neighborhood (b), 26-connected neighborhood (c) and our method (d). The RMS values of the surfaces are (b) 0.024, (c) 0.025 and (d) 0.022. The grid resolution is $60\times 60\times 60$.

We also compared the reconstruction of a horse model obtained using our continuous method with the reconstructions obtained using the discrete graph-cut method [16]. The results of our experiments are shown in Figure 8 and the complexity and accuracy of the reconstructions are described in Table 3. The results demonstrate that the continuous global optimization method returns a more accurate reconstruction and runs in less memory requirements in comparison with results presented in [16].

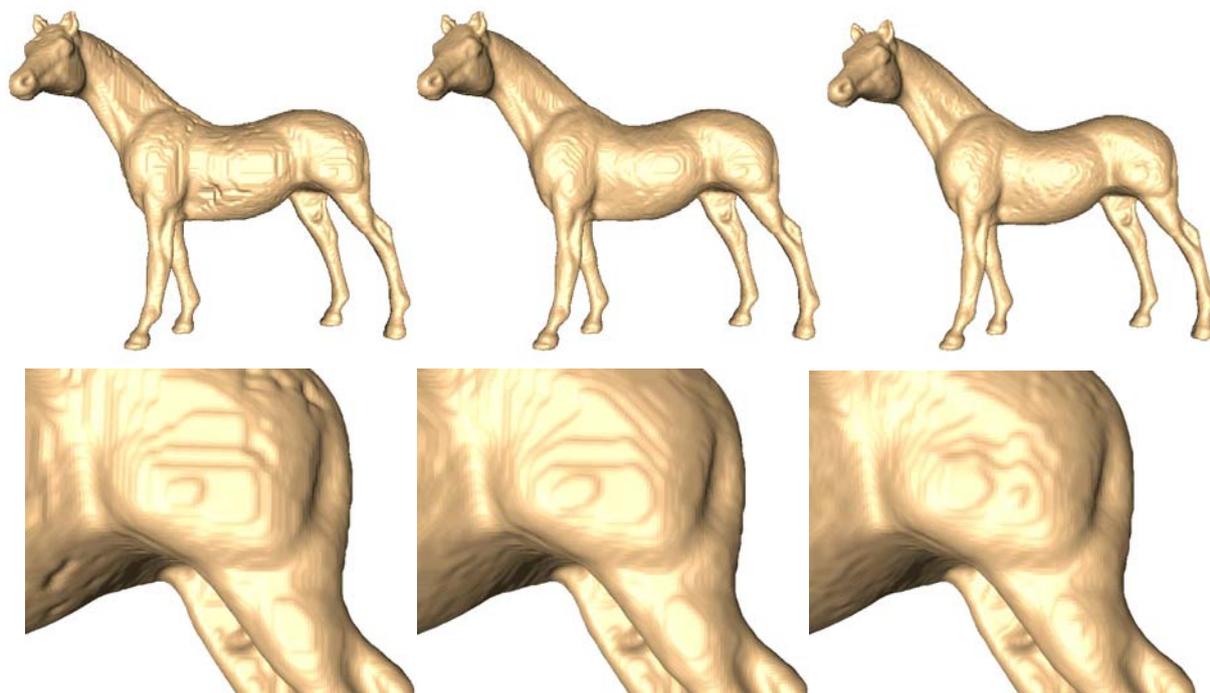

Figure 8: Reconstructed surfaces of a horse model from 100,000 sample points using discrete graph-cut algorithm of 6-connected neighborhood (left), 26-connected neighborhood (middle) and our continuous method (right). The bottom shows zoom-in of the models. The grid resolution is $196 \times 412 \times 345$.

| Method | Time[s] | Peak Memory[MB] | RMS | Triangles |
|---|---|---|---|---|
| Graph-cut of 6-neighborhood | 491 | 490 | 1.03e-4 | 490,332 |
| Graph-cut of 26-neighborhood | 548 | 1,432 | 1.01e-4 | 494,920 |
| Ours | 299 | 212 | 0.82e-4 | 498,220 |

Table 3: A comparison of the reconstruction time, the peak memory usage, the accuracy and the number of triangles in the reconstructed models using discrete graph-cut algorithm of 6-connected neighborhood, 26-connected neighborhood and our continuous method.

Regarding the memory consumption, the continuous optimization method clearly outperforms the discrete graph-cut methods. It requires only one floating point value for each voxel. In contrast, graph-cut methods require storage of edges and one flow value for each edge. This prohibits the usage of graph-cut optimization for high resolutions. With respect to the computation time, the continuous optimization method is slower than the graph-cut based approach as our current implementation is based on the SOR method on CPU and at least 200 iterations are required to get a reasonable result. Number if iterations can be decreased using better iterative method to solve Eq.(8).

However, the total variation method is inherently suitable for parallel computing and allows for a speed up factor of about 20 on GPU compared to the CPU version according to the results in [19]. The presented approach is convenient also for new CPU architectures, e.g.Intel's Single-Chip Cloud Computer (SCC) architecture.

# 6 Conclusion

We have presented a continuous global optimization approach for surface reconstruction from point clouds. The method is **robust** to noise, large holes and non-uniform sampling density under the condition of very coarse orientation information. As the proposed method is based on global approach and is inheritably reliable for different data sets it is slower than specialized method, similarly as the local and global optimization methods.

There are several future works to pursue. Because of discretizing on a regular 3D grid, the proposed method becomes slower for very fine-detailed reconstruction, nevertheless the advantages of the presented approach justify the computational expenses. In the case of practical problems solution the reliability and robustness are much more important than actual computations time, if acceptable, as the presented approach is not intended for "real-time" computation and rendering.

In the future, we intend to use an adaptive data structures. Accelerating the computation using GPU or other hardware platforms supporting parallel processing is another point of interest.

## Acknowledgments

The author would like to thank Victor Lempitsky (University of Oxford) and Kalin Kolev (Technical University of Munich) for useful discussions. This work was supported by the Key Project in the National Science & Technology Pillar Program of China (Grant No. 2008BAH29B02), the Shandong Natural Science Foundation of China (Grant No. ZR2010FM046), projects of the Ministry of Education of the Czech Republic No. 2C06002 and ME10060.

**Comments on Eq 8**

I do not know, how the implementation is actually made – sorry for the "stupid" comments.

- In which part do you spent majority computational time? – have you used a profiler?
- I would remove lambda multiplication
  $$\ldots\ldots\ \omega \frac{\sum + \sum + div\ v/\lambda}{\sum}$$
- If I understand the implementation you have to
  - Find all the neignbors for the $U_j$
- You probably use 3D array to store values – it is extremely expensive as there is no data coherence, swapping, mapping function T[i,j,k] -> Q[ii] is computed for each element ( i.e. 2x * +2*+ operations). Perhaps if one dimensional data structure is used, the evaluation is much faster.
- Selection of j from N(i) is time consuming, I expect . we usually use a simple mapping as we know that the neighbors are +1/-1/+N+1/+N-1/… and that is constant for the whole model.
  Note that it grows with $O(M^3)$ + some inner data structures

I expect that the timing is without reading/storing and rendering data.

Looking at Fig.1 I realized that there are many empty "cubicals". What about to remove those that are outside – we do not need to compute them – might be a significant speed up – can be solved just a simple bit map showing that the cube should not be computed.